\documentclass[11pt,onecolumn]{IEEEtran}

\def\BibTeX{{\rmfamily B\kern-.05em{\scshape i\kern-.025em b}\kern-.08em \TeX}}
\usepackage{epsfig, latexsym, cite, amsmath, graphicx}

\usepackage{amsthm}

\newtheorem{theorem}{Theorem}
\newtheorem{definition}{Definition}
\newtheorem{lemma}{Lemma}
\newtheorem{proposition}{Proposition}
\newtheorem{corollary}{Corollary}

\begin{document}

\title{Route Discovery and Capacity of Ad hoc Networks}

\author{Eugene Perevalov, Rick S. Blum, Xun Chen \and Anthony Nigara\\
Lehigh University,
Bethlehem, PA 18015
}

\maketitle 

\begin{abstract}
Throughput capacity of large ad hoc networks has been shown to
scale adversely with the size of network $n$. However the need for
the nodes to find or repair routes has not been analyzed in this
context. In this paper, we explicitly take route discovery into
account and obtain the scaling law for the throughput capacity
under general assumptions on the network environment, node
behavior, and the quality of route discovery algorithms. We also
discuss a number of possible scenarios and show that the need for
route discovery may change the scaling for the throughput capacity
dramatically.

\begin{flushleft}
\end{flushleft}
\end{abstract}

\section{Introduction}
The subject of this paper is the effect of the route discovery
process (\textit{RDP}) on the throughput capacity of ad hoc
networks. Previous results for the network capacity and
throughput, like those found in \cite{CWN}-\cite{TG} (see also
\cite{GT}-\cite{ElGam} for an analysis of the effect of mobility
on throughput), ignore the route discovery process and focus
solely on the data traffic that ad hoc networks can support. On
the other hand, under certain conditions (nodes leaving and
joining, for example) the route discovery process can consume a
significant portion of network resources and become detrimental to
overall network performance and stability. For example, if more
route discovery processes are initiated than can be sustained,
then they will likely fail resulting in more retransmissions. In
this scenario, the network can become inundated with route request
({\it RREQ}) packets and the overall network throughput can
significantly decrease.

In the following, we determine the impact of the route discovery
process on network throughput by determining the asymptotic
behavior and scalability with the number of nodes for a network
that has both data and \textit{RDP} transmissions. Let $W$ be the
number of bits that a node can successfully transmit per unit
time. We characterize the throughput in terms of two additional
basic \textit{RDP} related quantities:
\begin{itemize}
\item The average time that a route stays intact once established:
$\tau(n)$. \item The function $G(\cdot)$ defined by $Q=G(f)$,
where $Q$ is the probability that an \textit{RDP} is successful,
and $f$ is the fraction of all nodes reached by that \textit{RDP}.
\end{itemize}

We show that two qualitatively different situations can be
distinguished.
\begin{enumerate}
\item $\tau(n)G\left({1\over n}\right)=o\left({1\over \sqrt{n\log
n}}\right)$. In this case, the \textit{RDP} resource usage is
severe enough to become the throughput bottleneck and change its
scaling compared to the case when all routes are known. The
throughput scales as
$${\cal T}(n)=\Theta\left({W\tau(n)G\left({1\over n}\right)} \right).$$
\item $\tau(n)G\left({1\over n}\right)=\Omega\left({1\over
\sqrt{n\log n}}\right)$. In this case the \textit{RDP} does not
affect the throughput significantly (in the order of magnitude
sense) and the main limiting factor for the throughput is still
the interference between data transmissions:
$${\cal T}(n)=\Theta\left({W\over \sqrt{n \log n}} \right).$$
\end{enumerate}

We apply these general results to some canonical examples, with
some specific but reasonable assumed models for $ \tau(n) $ and $
G\left({1\over n}\right) $, to first show that the actual scaling
of the throughput can be changed from the case where routing is
ignored. In fact, for two of these cases we show
$${\cal T}(n)=\Theta\left({W\over n} \right)$$
which implies routing can cause even more severe throughput
scaling problems in ad hoc networks.  This occurs, for example,
when new nodes join a network for which $ \tau(n) $ is assumed
constant with $n$.  On the other hand, later examples indicate
that extremely efficient route repair can lessen, and maybe even
eliminate, the just mentioned additional scaling problems.

\section{System Model}
\label{model}

We consider a wireless ad hoc network with $n$ nodes distributed
uniformly over a unit square area.  A node $i$ can be in two
states: state $D$ and state $N$. In the state $D$, it can transmit
data to its current destination $d(i)$, and in state $N$ it can't
transmit (for example, due to lack of route knowledge). Let us
assume that
\begin{itemize}
\item  Nodes constantly change their states from $D$ to $N$ and
back.

\item The length of a period during which a node stays in the
state $D$ has an expected value of $\tau(n)$ which is assumed to
be determined exogenously.

\item The length of a period during which a node stays in the
state $N$ has an expected value of $\xi(n)$ which is to be
determined in the course of analysis.
\end{itemize}
The identity of the destination $d(i)$ for the node $i$ may change
from one active period (when the node is in state $D$) to another.
In the case that the identity of $d(i)$ changes, we assume that
the new destination is selected randomly. We also assume that one
node can't play the role of a destination for two different source
nodes simultaneously.

When a node is in $N$ state, it tries to discover the route to its
new destination. For that purpose it broadcasts \textit{RREQ}
packets. Let $S_{RREQ}$ be the size (in bits) of a \textit{RREQ}
packet. Recall $W$ is the number of bits that a node can
successfully transmit per unit time. This implies that the
transmission of a \textit{RREQ} packet can be effected in a time
of
\begin{equation}
\delta t= {S_{RREQ}\over W}. \label{eq:delt}
\end{equation}
In the following we assume that all time is slotted with the slot
size equal to $\delta t$. In any time slot a node can either
(re)transmit a data packet of size equal to $S_{RREQ}$ or
(re)broadcast a \textit{RREQ} packet.

If in a given time slot all nodes are transmitting data packets
then transmission success is governed by the Protocol Model in
which the transmission from node $i$ to node $j$ within distance
of $r$ from $i$ is successful if and only if there is no other
transmitting node $k$ within the distance of $(1+\Delta)r$ from
$j$. Here $r$ is the common transmission range which cannot be
less than $\sqrt{\log n \over \pi n}$ to ensure that the network
is connected with high probability \cite{GK0}.

If all nodes are (re)transmitting \textit{RREQ} packets in a given
time slot, we use the following model for capture success.

A node can be in either listening mode or transmitting mode in any
time slot. A transmitting node may either be initiating its own
\textit{RDP} or forwarding a \textit{RREQ} packet for another
node's \textit{RDP}. A node will be in the listening mode only
when it has no \textit{RREQ} packets to broadcast. If a node has
received a \textit{RREQ} in the previous time slot devoted to
\textit{RDP}, it will broadcast it in the next such time slot. A
node will not forward the same \textit{RREQ} packet more than
once. It is assumed that all nodes use the same transmission power
for broadcasting and a transmission by a node can be received in a
region of area $a$ around the node. Here $a\le 1$, and we assume
that $a(n)=\Theta\left({1\over n}\right)$ to model the situation
in which the physical transmission range of a node does not depend
on the number of nodes in the network\footnote{Indeed, if $a_{ph}$
is a physical transmission range, then, since we keep the system
area equal to 1, $a(n)={a_{ph}\over A(n)}$, where $A(n)$ is the
system area in physical units. The statement now follows when we
realize that $A(n)=\Theta(n)$.}.

A node will successfully receive a {\it RREQ} packet if the {\it
SINR} for that packet is larger than all others it receives during
the same time slot. Suppose there are $n_t$ nodes in transmitting
mode in the current time slot, denoted by $t_i$,
$i=1,2,\cdots,n_t$. Then there are $n-n_{t}=n_r$ nodes in
listening mode in the network, which we denote by $s_j$,
$j=1,2,\cdots,n_r$. For node $s_j$, there are only 3 possible
cases: i) if it is out of the transmission range of any $t_i$, it
will not receive anything, and it will stay in listening mode
until the next time slot; ii) if it is only within the
transmission range of one $t_i$, it will receive the \textit{RREQ}
packet from $t_i$; iii) if it is within the transmission range of
multiple $t_i$'s, node $s_j$ will successfully receive the
\textit{RREQ} packet that has the highest \textit{SINR} as
described in the capture model.

For the goal of this paper, we do not need to consider the
situation when both data and \textit{RREQ} packet transmissions
take place in the same time slot. A time multiplexing approach to
be described later is employed.

We assume that a node in state $N$ attempts to find a route. For
this purpose, it initiates an \textit{RDP} at a rate $\nu$. In the
following we will measure both time periods and rates in time
slots of physical length $\delta t$ with $\delta t$ given by
(\ref{eq:delt}). This implies that the rate of \textit{RDP}
initiation by any node does not exceed~1:
$$\nu\le 1.$$

\section{\textit{RDP} Success Probability}

The key measure of the effectiveness of a route discovery process
is the probability that it succeeds in finding the route. So we
have to be able to characterize the  probability of success of an
\textit{RDP} in the given environment. We will do it using the
following definition.
\begin{definition}
 Let $G(\cdot)$ be a monotonically
increasing function on the interval $[0,1]$ such that $G(0)=0$ and
$G(1)=1$. Then, if $f_k$ is the fraction of nodes that the
\textit{RDP} process $k$ has reached, the probability of a
successful route discovery by the process $k$ is $Q_k=G(f_k)$.
\end{definition}

When thinking of possible shapes of the function $G(\cdot)$, it is
reasonable to assume that the \textit{RDP} processes are ``totally
random'' in the worst case. In other words, it is reasonable to
exclude cases in which the probability of a node finding its
destination is lower than the fraction of all nodes reached by the
corresponding \textit{RDP} process. The latter situation is in
principle possible. For example consider the situation in which
the new nodes join the network in locations that are correlated
with the locations of the corresponding destinations. If the
correlation is such that the average distance between the source
and destination exceeds the average distance in the network, it is
possible to have $G(f)<f$ for $0<f<1$. However it is fairly clear
that such a situation is ``unnatural'' and we assume that nothing
like this actually happens. With this assumption, we have the
following.

\begin{proposition}
 $G(f)\ge f$ for $0\le f \le 1$.
\end{proposition}

Since, clearly, $G(0)=0$ and $G(1)=1$, it is also reasonable to
assume that the function $G(f)$ is concave.

We also introduce an unconditional route discovery success
probability $Q$ which we will treat as a function of the total
\textit{RDP} arrival rate in the network $\lambda$. It turns out
to be possible to put an upper bound on $Q$ in terms of $\lambda$
and the average number $\bar n_r$ of first-time \textit{RREQ}
packet receptions in a time slot. The proof of the next lemma is
omitted due to lack of space.

\begin{lemma}
If the route discovery is described by the function $G(f)$, then
the unconditional route discovery probability $Q$ is upper bounded
as
$$Q\le G(\bar f)=G\left({\bar n_r\over \lambda (n-1)}\right).$$
\label{l:Qbound}
\end{lemma}

\section{Network Capacity when Nodes can't always Transmit}
First, let us consider the case when, for large $n$, the average
length of active periods (when nodes are in the $D$ state) is not
much smaller than that of period of ``dormancy'' (when nodes are
in the $N$ state). In the asymptotic notation, this means that
$$\tau(n) = \Omega(\xi(n)).$$ In this case, it is easy to see that
the results on capacity reported in \cite{CWN} are valid.

Next, consider the case when the average length of active periods
becomes negligible compared to the ``dormant'' ones as the network
size $n$ increases, i.e.
$$\tau(n)=o(\xi(n))$$
in the asymptotic notation. For this case we have a different
upper bound on the per node throughput of the network. To find it,
we need an auxiliary result stated as a lemma whose proof is quite
straightforward (see \cite{rdp_cap}) and is omitted here.

\begin{lemma}
The proportion of time spent by a node in the state $D$ is
$${\tau(n)\over \tau(n) + \xi(n)}.$$
\label{l:active}
\end{lemma}

With the above lemma, we can obtain a ``dormancy induced'' upper
bound on the throughput.
\begin{theorem}
If every node alternates between states $D$ and $N$ spending an
average of $\tau(n)$ time slots in state $D$ and an average of
$\xi(n)$ time slots in state $N$ then the per node throughput is
$${\cal T}(n)=O\left( {W\tau(n)\over \xi(n)}\right)$$
\label{th:T_ub1}
\end{theorem}
\begin{proof}
Consider long time $T$ (measured in \textit{RDP} time slots).
According to Lemma~\ref{l:active}, only $T{\tau(n)\over
\tau(n)+\xi(n)}$ of these time slots can be used by any node for
data transmission. During these time slots a node can send no at
most $T{\tau(n)\over \tau(n)+\xi(n)}S_{RREQ}$ bits to its
destinations. So the inequality
\begin{equation}
{\cal T}(n) T \delta t \le T{\tau(n)\over \tau(n)+\xi(n)}S_{RREQ}
\label{eq:ineq_th}
\end{equation}
has to hold. Since $\delta t = {S_{RREQ}\over W}$, we obtain from
(\ref{eq:ineq_th}) that
$${\cal T}(n)\le {W\tau(n)\over \tau(n)+\xi(n)}\le {W\tau(n)\over
\xi(n)},$$ which proves the theorem.
\end{proof}

On the other hand, regardless of states of nodes, we have the
following upper bound on the throughput induced by interference
between simultaneous data transmissions.
\begin{theorem}
The per node throughput ${\cal T}(n)$ is upper bounded as
$${\cal T}(n) = O\left( {W\over \sqrt{n\log n}}\right). $$
\label{th:T_ub2}
\end{theorem}
The proof is standard in the literature on capacity of ad hoc
networks and is omitted.

Combining Theorems~\ref{th:T_ub1} and \ref{th:T_ub2}, and choosing
the tighter bound depending on the behavior of the ratio
${\tau(n)\over \xi(n)}$ we obtain the following corollary.

\begin{corollary}
The per node throughput ${\cal T}(n)$ is upper bounded as
$${\cal T}(n) = O\left({W\tau(n)\over \xi(n)}\right) $$
if ${\tau(n)\over \xi(n)}=o(1/\sqrt{n\log n})$ and it is upper
bounded as
$${\cal T}(n) = O\left({W\over \sqrt n}\right) $$
if ${\tau(n)\over \xi(n)}=\Omega(1/\sqrt{n\log n}))$.
\label{c:T_ub}
\end{corollary}

In order to show that the bounds of Corollary~\ref{c:T_ub} are
achievable up to constant we will demonstrate that there exists a
feasible transmission schedule that allows us to obtain the
required per node throughput. To achieve that goal, we tesselate
the square region into identical square cells with side of size
\begin{equation}
g(n)=\sqrt{2\log n\over n}. \label{eq:size}
\end{equation}

Two cells are called interfering neighbors if there is a point in
one cell within a distance of $(2+\Delta)r(n)$ from a point in the
other cell. It is easy to see that only transmissions from the
cells that are interfering neighbors can interfere with each
other. The following two lemmas are standard in the literature on
ad hoc network capacity (see for example \cite{CWN}).

\begin{lemma}
There exists a transmission schedule in which each cell can
transmit in one of every $\tilde c + 1$ time slots, where $\tilde
c$ depends only on the parameter $\Delta$. \label{l:sched}
\end{lemma}

\begin{lemma}
The probability that there is a cell that does not contain a
single node is upper bounded by
\begin{equation}
{1 \over 2 n\log n} , \label{lemma5}
\end{equation}
In other words, all cells contain at least one node with high
probability.
\end{lemma}

We organize transmission in the following way. The entire system
is tesselated into square cells of area ${g(\rho_s)}^2$ .

The routing of packet between nodes proceeds as follows. To route
a packet between two nodes, we employ at most two straight lines:
one vertical and one horizontal\footnote{It is possible that only
one straight line is needed}. Each time a packet is transmitted
from a node in a cell to some node in an adjacent cell. In the
final hop, the packet is transmitted to the destination from a
node in a cell adjacent to the cell containing the destination.

Now, let us consider a given cell $C_i$ and count the number of
routes passing through it. Let us denote this number by $N_i$. For
$N_i$, we have the following two results the proof of which makes
use of Chernoff bound and is fairly straightforward. It can be
found in \cite{rdp_cap} and is omitted here due to size
limitations.
\begin{lemma}
$$\max_i N_i =O\left( \sqrt{n \log n}\right)$$
with high probability. \label{l:maxNi}
\end{lemma}

\begin{lemma}
$$\min_i N_i = \Omega(\sqrt{n \log n})$$
with high probability.
\end{lemma}

We can now find the achievable per node throughput. This is the
subject of the next two theorems.
\begin{theorem}
If $${\tau(n)\over \xi(n)}=o\left({1\over \sqrt{n\log
n}}\right),$$ then the per node throughput
$${\cal T}(n) = \Omega\left({W\tau(n)\over \xi(n)}\right)$$
is achievable with high probability. \label{th:T_lb1}
\end{theorem}
\begin{proof}
Consider a long time $T$. Since each source can generate data only
in the $D$ state, using Lemmas \ref{l:active} and \ref{l:maxNi},
we see that the number of packets $N_{T,i}$ that has to be served
by the cell $C_i$ can be upper bounded as
$$N_{T,i}\le
\max_i N_i {\tau(n)\over \xi(n)}T=c\sqrt{n\log n} {\tau(n)\over
\xi(n)}T$$ with high probability. Since ${\tau(n)\over
\xi(n)}=o\left({1\over \sqrt{n\log n}}\right)$, we have that, with
high probability,
$$N_{T,i} = o(T),$$
which is less than the number of time slots ${T\over \tilde c}$
that, as shown in lemma \ref{l:sched} each cell can be active in.
This implies that the per node throughput of
$${{\tau(n)\over \xi(n)}S_{RREQ} T\over T\delta t} = {\tau(n)\over
\xi(n)}W$$ is achievable with high probability, which proves the
theorem.
\end{proof}
The meaning of the next theorem is that, if the ratio
$\tau(n)\over \xi(n)$ is large enough, the throughput limited by
the interference between data transmissions can be achieved. Its
proof is similar to that of the previous theorem and is omitted.

\begin{theorem}
If $${\tau(n)\over \xi(n)}=\Omega\left({1\over \sqrt{n\log
n}}\right),$$ then the per node throughput
$${\cal T}(n) = \Omega\left({W\over \sqrt{n\log n}}\right)$$
is achievable with high probability. \label{th:T_lb2}
\end{theorem}

\section{Scaling of $\xi(n)$}
A node that needs to find a route will initiate an \textit{RDP}.
Since it may not be successful, the node might have to initiate it
several times. We assume that the node initiates \textit{RDP}'s
with frequency of $\nu$ until the route is found. The next lemma
computes the expected number of \textit{RDP}'s that a node will
need to initiate in order to find the route. The proof of the
following lemma is elementary.

\begin{lemma}
The average number of \textit{RREQ} transmissions, $N_{avg}$,
which is required by a node for a successful route discovery to
the same destination, is given by
\[N_{avg}=\frac{1}{Q}\]
\label{Navg}
\end{lemma}

Let us introduce $Q'(\lambda)$ as the unconditional probability
for a success of an \textit{RDP} in the presence of data
transmission. Obviously, $Q'(\lambda)$ may depend on how the
network resources (time and space) are divided between
\textit{RDP} and data transmission. It's clear that for any such
scheme
\begin{equation}
Q'(\lambda)\le Q(\lambda)\label{eq:Q_ineq}.
\end{equation}

The next lemma relates the frequency of \textit{RDP} initiation
$\nu$, the average time of route life $\tau(n)$ and the total
\textit{RDP} arrival rate $\lambda$.

\begin{lemma}\label{new_lmdeqn}
The total \textit{RDP} arrival rate can be computed as
$$\lambda={n\nu \over \nu \tau(n) Q'(\lambda) +1}$$
\end{lemma}
\begin{proof}
As shown in Lemma~\ref{Navg}, a node needs to initiate on average
$\frac{1}{Q'(\lambda)}$ \textit{RDP}'s  for a successful route
discovery. The time between two consecutive initiations is given
by $\frac{1}{\nu}$. The total time of transmitting
$\frac{1}{Q'(\lambda)}$ \textit{RREQ} packets is then
$\frac{1}{Q'(\lambda)\nu}$. After this total time, the valid route
is set up, and lasts for time $\tau(n)$ during which the node can
transmit data and no route discovery is required.

Thus, on average, each node initiates $\frac{1}{Q'(\lambda)}$
\textit{RREQ} packets every time period of
$\frac{1}{Q'(\lambda)\nu}+\tau(n)$. Therefore the \textit{RDP}
arrival rate due to one node can be found as
\begin{equation}
\frac{\lambda}{n} = \frac{\frac{1}{Q'(\lambda)}}{\frac{1}{Q\nu}+\tau(n)} \nonumber \\
 = \frac{\nu}{1+Q'(\lambda)\tau(n)\nu},
\end{equation}
yielding,
\begin{equation}
\label{reallmd} \lambda=\frac{n\nu}{1+Q'(\lambda)\tau(n)\nu}
\end{equation}
for the total \textit{RDP} arrival rate.
\end{proof}

We would like to demonstrate that the average length of a node
``inactivity'' period (when it is in the state $N$ attempting to
discover a route) is bounded from below and the bound depends on
the shape of the route discovery success function $G(\cdot)$.

We can now state the lower bound result.
\begin{theorem}
The expected length of the time interval during which a node stays
in the $N$ state is
$$\xi(n)=\Omega\left({1\over G\left({1\over n}\right)}\right).$$
\label{th:xi_lb}
\end{theorem}
The proof makes use of Lemmas \ref{l:Qbound} and \ref{new_lmdeqn}
and is omitted due to lack of space.

Next, we would like to find an upper bound on the average length
of ``data inactivity'' period $\xi(n)$. Note that, in order to
find a lower bound, it was sufficient to assume that all network
resources were devoted to route discovery with no data
transmission taking place. For an upper bound, we need to present
a constructive network resource division scheme between
\textit{RDP} and data transmission.

We make the following assumption about the probability
distribution of the fraction of nodes reached by an \textit{RDP}.

\noindent{\bf Assumption 1:} The probability distribution of the
fraction of nodes reached by an \textit{RDP} is such that $m_f\ge
\gamma\bar f$, where $m_f$ is the median of the distribution, and
$\gamma>0$ is a constant independent on $n$.

Note that the goal of making this assumption is to rule out
``pathological'' distributions of the fraction of nodes reached by
an \textit{RDP}. Thus it is not restrictive in that any
distribution that can be realized in practice should satisfy it
for an appropriate value of $\gamma$.

Let us adopt a time division scheme in which in any given time
slot either only data transmission or only route discovery takes
place. In other words, all time slots are divided into two types:
data slots and \textit{RDP} slots. During data slots, only the
data transmission takes place whereas during \textit{RDP} slots
only \textit{RDP} processes are allowed to run. Let $\theta$ be
the fraction of \textit{RDP} time slots. We call this resource
division scheme ``Scheme A''.

 Let us denote by $Q'$ the success
probability in the presence of data transmission. In the time
division scheme that we are using, it is straightforward to find
$Q'(\lambda)$.

\begin{lemma}
Under Scheme A, the unconditional probability of route discovery
success $Q'(\lambda)$ can be found as $Q'(\lambda)=Q({\lambda\over
\theta})$.
\end{lemma}

With the new success probability $Q'(\lambda)$ is is
straightforward to find the relation between the rate of retrial
for every node $\nu$ and the total arrival rate $\lambda$. For
this we just need to repeat the steps that led to the statement of
lemma \ref{new_lmdeqn} but with $Q'$ instead of $Q$. We state the
result as a lemma.

\begin{lemma} Under the time division Scheme A, the following relation holds
between $\nu$ and $\lambda$.
$$\lambda = {n\nu \over \nu \tau Q'(\lambda) +1}.$$
\label{l:lambda_Q'}
\end{lemma}

To derive an upper bound on the average length $\xi(n)$ of ``data
inactivity'' period, we need the following auxiliary results whose
proof is omitted due to lack of space.

\begin{lemma}
$${\bar n_r\over n}\ge \hat c,$$
where $\hat c$ is a constant independent on $n$. \label{l:n_r_lb}
\end{lemma}

The next lemma puts a lower bound on the unconditional probability
of success of an \textit{RDP} under Assumption 1.

\begin{lemma}
If Assumption 1 holds, then $Q(\lambda)\ge {1\over
2}G\left({\gamma \bar n_r\over \lambda n}\right)$. \label{l:Q_lb}
\end{lemma}
\begin{proof}
Since $Q(\lambda)=\overline{G(f)}$, by definition of a median, we
have
$$Q(\lambda)=\overline{G(f)} \ge {1\over 2}G(m_f)\ge {1\over 2}G(\gamma
\bar f)={1\over 2}G\left({\gamma \bar n_r\over \lambda
n}\right),$$ which proves the lemma.
\end{proof}

We can now use the above lemmas to find an upper bound on
$\xi(n)$.

\begin{theorem}
Under scheme A
$$\xi(n)= O\left( {1\over G\left({1\over n }\right)}\right).$$
\label{th:xi_ub}
\end{theorem}
The proof of this theorem follows in a straightforward fashion
from Lemmas \ref{l:n_r_lb} and \ref{l:Q_lb} and is omitted due to
size limitations.

Putting together the results of Theorems~\ref{th:xi_lb} and
\ref{th:xi_ub}, we obtain a tight asymptotic characterization of
the quantity $\xi(n)$ which we state as a corollary.

\begin{corollary}
$$\xi(n)=\Theta\left({1\over G\left({1\over n} \right)} \right).$$
\label{c:xi}
\end{corollary}

\section{\textit{RDP} limited
throughput} In this section, we collect the pieces to obtain the
main result of this paper: the scaling of the \textit{RDP} limited
throughput of a random ad hoc network. The next theorem covers the
case where the \textit{RDP} plays the role of the throughput
bottleneck.

\begin{theorem}
If $\tau(n) G\left({1\over n}\right)=o\left({1\over \sqrt{n\log
n}}\right)$ then
$${\cal T}(n)=\Theta\left(W\tau(n) G\left({1\over n}\right) \right). $$
\label{th:main1}
\end{theorem}
\begin{proof}
If $\tau(n) G\left({1\over n}\right)=o\left({1\over \sqrt{n\log
n}}\right)$ then, according to Corollary~\ref{c:xi},
$${\tau(n)\over \xi(n)} = o\left({1\over \sqrt{n\log
n}}\right).$$

On the other hand, combining the results of
Theorems~\ref{th:T_ub1} and \ref{th:T_lb1}, we obtain that if
${\tau(n)\over \xi(n)} = o\left({1\over \sqrt{n\log n}}\right)$,
then
$${\cal T}(n)=\Theta\left({W\tau(n) \over \xi(n)}\right) $$
 and therefore (Corollary~\ref{c:xi})
$${\cal T}(n)=\Theta\left(W\tau(n) G\left({1\over n}\right)
\right),$$ which proves the theorem.
\end{proof}

The following theorem shows that when the product $\tau(n)
G\left({1\over n}\right)$ is large enough, the main limiting
factor for the throughput is the interference between simultaneous
data transmissions and we are back to the case described in
\cite{CWN}.

\begin{theorem}
If $\tau(n) G\left({1\over n}\right)=\Omega\left({1\over
\sqrt{n\log n}}\right)$ then
$${\cal T}(n)=\Theta\left({W\over \sqrt{n\log n}} \right). $$
\label{th:main2}
\end{theorem}
\begin{proof}
If $\tau(n) G\left({1\over n}\right)=\Omega\left({1\over
\sqrt{n\log n}}\right)$ then, according to Corollary~\ref{c:xi},
$${\tau(n)\over \xi(n)} = \Omega\left({1\over \sqrt{n\log
n}}\right).$$

Now, combining the results of Theorems~\ref{th:T_ub2} and
\ref{th:T_lb2}, we see that if ${\tau(n)\over \xi(n)} =
\Omega\left({1\over \sqrt{n\log n}}\right)$, then
$${\cal T}(n)=\Theta\left({W\over \sqrt{n\log n}}\right), $$
which proves the theorem.
\end{proof}

Now that the general asymptotic behavior of the \textit{RDP}
limited throughput is given, let us consider several simple
examples and obtain the throughput scaling.

\underline{Examples:}

1. The need of route discovery due to changing node membership. In
this model, a new pair of nodes $i$ and $j$ join the network and
need to discover a route to each other. They stay connected during
a time period of $\tau_{ij}$ after which the connection is
terminated and the nodes leave the network (turn themselves off).
This process of pairs joining the network and leaving continues is
such a way that the network is in an equilibrium in the sense of
the number of nodes participating in it at any given time. Let us
assume that $E(\tau_{ij})=\tau$ for any $i$ and $j$. In this case
it is reasonable to assume that $\tau$ does not depend on $n$, as
it depends only on the behavior pattern of individual nodes. Since
both nodes that just have joined the network are ``new'' to it,
the \textit{RDP} success probability function is $G(f)=f$ . This
implies that
$$G\left({1\over n}\right)=\Theta\left({1\over n}\right),$$
and therefore,
$$\tau(n)G\left({1\over n}\right)=\Theta\left({1\over
n}\right)=o\left({1\over \sqrt{n\log n}} \right).$$ Then it
follows form Theorem~\ref{th:main1} that the \textit{RDP} limited
throughput scales as
$${\cal T}(n)=\Theta\left({W\over n} \right).$$

2. The node membership in the network is constant but the loss of
routes is due to severe fading or excessive node mobility. In this
case, it would be reasonable to assume that a route from a node
$i$ to $d(i)$ is lost and has to be rediscovered whenever one (or
another constant number) link is broken. The number of links
between $i$ and $d(i)$ is $\Theta\left({1\over \sqrt n}\right)$,
and the rate at which a link brakes depends only on fading
environment and the mobility characteristics and therefore is
independent on $n$. Therefore, assuming that the links brake
independently, we obtain that the rate at which the nodes $i$ and
$d(i)$ lose their route will be $\Theta\left({1\over \sqrt
n}\right)$.  Hence, the average length of time during which the
route stays intact is
$$\tau(n)=\Theta\left({1\over \sqrt n} \right).$$
On the other hand, since both the source and the destination are
still present in the network, they are ``known'' to $\Theta(\sqrt
n)$ other nodes (the number of routes passing through a typical
node). This means that in order to discover a route to the
destination, an \textit{RDP} initiated by the source needs to
reach one of $\Theta(\sqrt n)$ nodes in the network. Therefore,
$$G\left({1\over n} \right)=\Theta\left({1\over \sqrt n}
\right),$$ and hence (Theorem~\ref{th:main1}),
$${\cal T}(n)=\Theta\left({W\over n} \right).$$

3. The situation is just as above with the exception of the route
discovery probability. Assume that the route repair algorithm is
so good that it is able to repair the broken link immediately so
that the CL model is appropriate. Then
$$\tau(n)=\Theta\left({1\over \sqrt n} \right),$$
as above, but
$$G\left({1\over n} \right)=\Theta\left(1\right).$$ In this case,
Theorem~\ref{th:main2} yields
$${\cal T}(n)=\Theta\left({W\over \sqrt{n\log n}} \right).$$
In other words, the scaling is the same as in the case with no
route discovery meaning that under such conditions the main
limitation is still data transmission.

\end{document}